\documentclass[aps,preprint,showpacs]{revtex4-1}
\usepackage{amsfonts}
\usepackage{amsmath}
\usepackage{amssymb}
\usepackage{graphicx}
\begin{document}

\title{Corroborating the equivalence between the Duffin-Kemmer-Petiau and the
Klein-Gordon and Proca equations}
\author{L. B. Castro}\email[ ]{lrb.castro@ufma.br}
\affiliation{Departamento de F\'{\i}sica CFM,
Universidade Federal de Santa Catarina (UFSC), 88.040-900  CP. 476, Florian\'{o}polis, Santa Catarina, Brazil}
\affiliation{Departamento de F\'{\i}sica, Universidade Federal do Maranh\~{a}o (UFMA), Campus Universit\'{a}rio do Bacanga, 65085-580, S\~{a}o Lu\'{i}s, Maranh\~{a}o, Brazil}
\author{A. S. de Castro}\email[ ]{castro@pq.cnpq.br}
\affiliation{Departamento de F\'{\i}sica e Qu\'{\i}mica,
Universidade Estadual Paulista, 12516-410 Guaratinguet\'{a}, S\~ao
Paulo, Brazil}


\begin{abstract}
It is shown that the Hamiltonian version of the Duffin-Kemmer-Petiau theory
with electromagnetic coupling brings about a source term at the current. It
is also shown that such a source term disappears from the scenario if one
uses the correct physical form for the Duffin-Kemmer-Petiau field, regardless the choice for representing the Duffin-Kemmer-Petiau matrices. This result is used to fix the ambiguity in the electromagnetic coupling in the Duffin-Kemmer-Petiau theory. Moreover, some widespread misconceptions about the Hermiticity in the Duffin-Kemmer-Petiau theory are discussed.
\end{abstract}

\maketitle

\section{Introduction}

The first-order Duffin-Kemmer-Petiau (DKP) formalism \cite{pet,kem}
describes spin-0 and spin-1 particles and enjoys a richness of couplings not capable of being expressed in the Klein-Gordon (KG) and Proca theories \cite{guer,vija}. The way for introducing the minimal coupling has been a subject of quite some debate. One can introduce the minimal coupling either at the equation of motion or at the Hamiltonian form of the DKP theory, and these ways seem not to be equivalent as already noted by Kemmer in his original work \cite{kem}. The main issue is that when the minimally coupled covariant form of DKP equation is written in Hamiltonian form, there appears an additional term which is called anomalous term and additionally the energy-momentum tensor is not conserved. Ghose \cite{gho1} suggested that one should introduce the minimal coupling at the Hamiltonian form of the DKP theory for avoiding the appearance of the anomalous term, and a conserved energy-momentum tensor appears as a bonus. Nowakowski \cite{now} and Lunardi et al. \cite{lun} showed that such an anomalous term disappears when the physical components of DKP field are selected. Struyve et al. \cite{stru} analyzed the ambiguity of introducing the minimal coupling and suggested that despite the nonconservation of the energy-momentum tensor, we should introduce the minimal coupling via the covariant form of the DKP equation, in order to obtain the minimally coupled KG theory. Therefore, there exists a discrepancy in how introduce the minimal coupling and it seems that this discrepancy has still not found a definitive conclusion.

The main purpose of the present paper is to clarify the ambiguity of the electromagnetic coupling in the DKP theory. To achieve this, the continuity equation for a charged boson minimally coupled to the electromagnetic field is analyzed by using both the equation of motion
and its Hamiltonian version. It is shown that the charge quadricurrent $J^{\mu }$ has a source term
when one uses the Hamiltonian version of the DKP theory. By using a proper set of operators \cite{FIS} whose algebraic properties make our conclusions independent of the choice for representing the DKP matrices, it is also shown
that such a source term disappears from the DKP theory if one uses the
correct physical components of the DKP spinor as prescribed in \cite{lun}. Therefore, it does not matter if one either put the electromagnetic coupling straight in the Hamiltonian or in  the equation of
motion, because the current is conserved in both versions of the DKP theory. In addition, some widespread misconceptions about the Hermiticity in the DKP theory diffused in the literature are discussed.

\section{Duffin-Kemmer-Petiau equation}

The DKP equation for a free charged boson is given by \cite{kem} (with units in
which $\hbar =c=1$)%
\begin{equation}
\left( i\beta ^{\mu }\partial _{\mu }-m\right) \psi =0  \label{dkp}
\end{equation}%
\noindent where the matrices $\beta ^{\mu }$\ satisfy the algebra $\beta
^{\mu }\beta ^{\nu }\beta ^{\lambda }+\beta ^{\lambda }\beta ^{\nu }\beta
^{\mu }=g^{\mu \nu }\beta ^{\lambda }+g^{\lambda \nu }\beta ^{\mu }$%
\noindent and the metric tensor is $g^{\mu \nu }=\,$diag$\,(1,-1,-1,-1)$.
That algebra generates a set of 126 independent matrices whose irreducible
representations are a trivial representation, a five-dimensional
representation describing the spin-0 particles and a ten-dimensional
representation associated to spin-1 particles \cite{qed}. The DKP spinor has an excess of components and the theory has to
be supplemented by an equation which allows to eliminate the redundant
components. That constraint equation is obtained by multiplying the DKP
equation by $1-\beta ^{0}\beta ^{0}$, namely%
\begin{equation}
i\beta ^{j}\beta ^{0}\beta ^{0}\partial _{j}\psi =m\left( 1-\beta ^{0}\beta
^{0}\right) \psi ,\quad j\text{ \ runs from 1 to 3}  \label{vin1}
\end{equation}%

\noindent This constraint equation expresses three (four) components of the spinor by
the other two (six) components and their space derivatives in the scalar
(vector) sector so that the superfluous components disappear and there only
remain the physical components of the DKP theory. The second-order KG and Proca equations are obtained when one
selects the spin-0 and spin-1 sectors of the DKP theory. The DKP theory has also
its Hamiltonian version in the form (see, e.g., \cite{now,lun})%
\begin{equation}
i\partial _{0}\psi =H\psi ,\quad H=i[\beta ^{j},\beta ^{0}]\partial
_{j}+m\beta ^{0}
\end{equation}%

\noindent Note that in this context we can show that $H^{\dagger}=H$. A well-known
conserved four-current is given by%
\begin{equation}
J^{\mu }=\frac{1}{2}\bar{\psi}\beta ^{\mu }\psi  \label{jota}
\end{equation}
\noindent\ where the adjoint spinor $\bar{\psi}$ is given by $\bar{\psi}%
=\psi ^{\dagger }\eta ^{0}$ with $\eta ^{0}=2\beta ^{0}\beta ^{0}-1$ in such
a way that $\left( \eta ^{0}\beta ^{\mu }\right) ^{\dagger }=\eta ^{0}\beta
^{\mu }$ (the matrices $\beta ^{\mu }$ are Hermitian with respect to $\eta
^{0}$). Despite the similarity to the Dirac equation, the DKP equation involves singular
matrices, the time component of $J^{\mu}$ given by~(\ref{jota}) is not
positive definite and the case of massless bosons can not be obtained by a
limiting process~\cite{nieto}. Nevertheless, the matrices $\beta^{\mu}$ plus
the unit operator generate a ring consistent with integer-spin algebra and $%
J^{0}$ may be interpreted as a charge density. The factor $1/2$ multiplying $\bar{\psi}\beta^{\mu}\psi$, of no importance regarding the conservation law, is in order to hand over a charge density conformable to that one used in the KG theory and its nonrelativistic limit \cite{ben1}.

\section{Interactions in the Duffin-Kemmer-Petiau equation}

With the introduction of interactions, the DKP equation can be written as
\begin{equation}
\left( i\beta ^{\mu }\partial _{\mu }-m-U\right) \psi =0  \label{dkp222}
\end{equation}

\noindent where the more general potential matrix $U$ is written in terms of
25 (100) linearly independent matrices pertinent to five (ten)-dimensional
irreducible representation associated to the scalar (vector) sector. In the
presence of interaction, $J^{\mu}$ satisfies the equation
\begin{equation}  \label{corrent222}
\partial _{\mu }J^{\mu }+\frac{i}{2}\,\bar{\psi}\left( U-\eta ^{0}U^{\dagger
}\eta ^{0}\right) \psi =0\,.
\end{equation}

\noindent Thus, if $U$ is Hermitian with respect to $\eta ^{0}$ then
four-current will be conserved. The potential matrix $U$ can be written in
terms of well-defined Lorentz structures. For the spin-0 (scalar sector) there are
two scalar, two vector and two tensor terms \cite{guer}, whereas for the
spin-1 (vector sector) there are two scalar, two vector, a pseudoscalar, two
pseudovector and eight tensor terms \cite{vija}. The tensor terms have been
avoided in applications because they furnish noncausal effects \cite{guer}-%
\cite{vija}. The condition (\ref{corrent222}) has been used to point out a misleading treatment in the recent literature regarding analytical solutions for nonminimal vector interactions \cite{ben4,ben5,ben2}.

\subsection{Duffin-Kemmer-Petiau equation with minimal electromagnetic coupling}

Considering only the minimal vector interaction, the DKP equation for a charged boson with minimal electromagnetic coupling is given
by%
\begin{equation}
\left( i\beta ^{\mu }D_{\mu }-m\right) \psi =0  \label{dkp2}
\end{equation}%
\noindent \noindent where the covariant derivative is given by $D_{\mu
}=\partial _{\mu }+ieA_{\mu }$. In this case, the constraint equation (\ref{vin1}) becomes
\begin{equation}\label{vin2}
i\beta ^{k}\beta ^{0}\beta ^{0}\partial _{k}\psi-e\beta ^{k}\beta ^{0}\beta ^{0}A_{k}\psi =m\left( 1-\beta ^{0}\beta
^{0}\right) \psi\,,
\end{equation}

\noindent and the four-current $J^{\mu}$ retains its form as (\ref{jota}). The DKP theory with minimal electromagnetic coupling has also its Hamiltonian
version in the form (see, e.g., \cite{now,lun})%
\begin{equation}
i\partial _{0}\psi =H\psi ,\quad H=i[\beta ^{j},\beta
^{0}]D_{j}+eA_{0}+m\beta ^{0}+\frac{ie}{2m}F_{\mu \nu }\left( \beta ^{\mu
}\beta ^{0}\beta ^{\nu }+\beta ^{\mu }g^{0\nu }\right)  \label{sch}
\end{equation}%

\noindent with the electromagnetic field tensor given by $F_{\mu \nu }=\partial _{\mu
}A_{\nu }-\partial _{\nu }A_{\mu }$. The last term in $H$, is called anomalous term because it has no equivalent in the spin-$1/2$ Dirac theory. For this reason it has been
suggested to put the electromagnetic coupling straight in the Hamiltonian instead of \ the equation
of motion \cite{gho1}. However, it has been shown in Refs~\cite{now,lun} that such an anomalous term disappears when the physical components of DKP field are selected. Since
\begin{equation}
\left( iF_{0j}\beta ^{j}\beta ^{0}\beta ^{0}\right) ^{\dagger }=-\left(
iF_{0j}\beta ^{j}\beta ^{0}\beta ^{0}\right) +iF_{0j}\beta ^{j}  \label{her}
\end{equation}%
$H$ is not equal to $H^{\dagger }$ \cite{tati}, in opposition what was adverted in \cite%
{now}. Because of this, the Lewis-Riesenfeld invariant method for studying
time-dependent fields is not straightforwardly applicable as done in Ref.
\cite{mer}, and already criticized in \cite{ben}.

\subsection{Hamilton form and Hermiticity}

At  this level, it is worthwhile to note that the Hamiltonian given by (\ref{sch}) should be Hermitian with respect to $\eta^{0}$ and not with respect a $\beta^{0}$ as was stated by Zeleny \cite{zel}. Zeleny argued that an operator, and in particular the Hamiltonian should be neo-Hermitian ($\beta^{0}\hat{O}=\hat{O}^{\dagger}\beta^{0}$). Furthermore, Zeleny claimed that the free Hamiltonian as well as the minimally coupled Hamiltonian are not neo-Hermitian. Nevertheless, it can be easily shown that both of them are Hermitian with respect to $\eta^{0}$, and therefore $\langle H\rangle$ is a real quantity.

\subsubsection{The free case}

\noindent The Hamiltonian form of the free DKP equation is given by
\begin{equation}\label{fhap}
    i\partial_{0}\psi=H\psi, \qquad H=i\left[ \beta^{j},\beta^{0} \right]\partial_{j}+m\beta^{0}.
\end{equation}

\noindent with the constraint equation
\begin{equation}\label{vin}
i\beta ^{k}\beta ^{0}\beta ^{0}\partial _{k}\psi =m\left( 1-\beta ^{0}\beta
^{0}\right) \psi \,.
\end{equation}

\noindent From the algebra of matrices $\beta^{\mu}$, it is shown that
\begin{equation}
    \left[ \beta^{i},\beta^{0} \right]^{\dagger}=\left[ \beta^{i},\beta^{0} \right]
\end{equation}

\noindent Furthermore $\hat{p}_{\mu}=i\partial_{\mu}$ is Hermitian and commutes with the matrices $\beta^{\mu}$. With all this, we can show that $H^{\dagger}=H$. On the other hand, multiplying (\ref{fhap}) by $\beta^{0}$ from the left and using the constraint (\ref{vin}), we obtain
\begin{equation}\label{betah}
    \beta^{0}H=-i\beta^{k}\partial_{k}+m
\end{equation}

\noindent Hermitian conjugation of (\ref{betah}), gives
\begin{equation}\label{fhher}
    \left( \beta^{0}H\right)^{\dagger}=i\beta^{k}\partial_{k}+m \neq \beta^{0}H
\end{equation}

\noindent from this result Zeleny \cite{zel} concluded that not even the free Hamiltonian form is neo-Hermitian, for this reason he tried to build a neo-Hermitian Hamiltonian. On the other hand, with the correct criteria it can be shown that
\begin{equation}\label{betaeta}
    \eta^{0}\left(\beta^{0}H\right)=-i\eta^{0}\beta^{k}\partial_{k}+\eta^{0}m
\end{equation}

\noindent and applying the Hermitian conjugation on (\ref{betaeta}) we have that
\begin{equation}\label{fhheret}
   \left[\eta^{0} \left( \beta^{0}H\right)\right]^{\dagger}=i\beta^{k}\eta^{0}\partial_{k}+\eta^{0}m =\eta^{0} \left(\beta^{0}H\right)\,.
\end{equation}

\noindent Therefore, the Hamiltonian form of the free DKP theory is Hermitian with respect to $\eta^{0}$.

\subsubsection{The electromagnetic case}

\noindent The Hamiltonian form for the minimally coupled case has the form
\begin{equation}\label{fhem}
    i\partial_{0}\psi=H\psi, \quad H=i\left[ \beta^{j},\beta^{0} \right]D_{j}+eA_{0}^{(1)}+m\beta^{0}+\frac{ie}{2m}\,F_{\mu\nu}\left( \beta^{\mu}\beta^{0}\beta^{\nu}+\beta^{\mu}\eta^{0\nu} \right)
\end{equation}

\noindent Furthermore, the constraint becomes
\begin{equation}\label{vin2}
i\beta ^{k}\beta ^{0}\beta ^{0}\partial _{k}\psi-e\beta ^{k}\beta ^{0}\beta ^{0}A_{k}\psi =m\left( 1-\beta ^{0}\beta
^{0}\right) \psi\,.
\end{equation}

\noindent Since
\begin{equation}
\left( iF_{0j}\beta ^{j}\beta ^{0}\beta ^{0}\right) ^{\dagger }=-\left(
iF_{0j}\beta ^{j}\beta ^{0}\beta ^{0}\right) +iF_{0j}\beta ^{j}
\end{equation}%

\noindent $H$ is not equal to $H^{\dagger}$. On the other hand, multiplying (\ref{fhem}) by $\beta^{0}$ from the left and using the constraint (\ref{vin2}), we obtain
\begin{equation}\label{betah2}
    \beta^{0}H=-i\beta^{k}\partial_{k}+e\beta^{k}A_{k}+e\beta^{0}A_{0}+m\,.
\end{equation}

\noindent Taking the Hermitian conjugation of (\ref{betah2}) we have
\begin{equation}\label{fhher}
    \left( \beta^{0}H\right)^{\dagger}=i\beta^{k}\partial_{k}-e\beta^{k}A_{k}+e\beta^{0}A_{0}+m \neq \beta^{0}H
\end{equation}

\noindent similarly to the free case. On the other hand, it can be shown that
\begin{equation}\label{betaeta2}
    \eta^{0}\left(\beta^{0}H\right)=-i\eta^{0}\beta^{k}\partial_{k}+e\eta^{0}\beta^{k}A_{k}+e\beta^{0}A_{0}+\eta^{0}m
\end{equation}

\noindent and applying the Hermitian conjugation in (\ref{betaeta2}), we obtain
\begin{equation}\label{fhheret2}
   \left[\eta^{0} \left( \beta^{0}H\right)\right]^{\dagger}=i\beta^{k}\eta^{0}\partial_{k}-e\beta^{k}\eta^{0}A_{k}+e\beta^{0}A_{0}+\eta^{0}m =\eta^{0} \left(\beta^{0}H\right)
\end{equation}

\noindent Therefore the Hamiltonian form with electromagnetic interaction of the DKP theory is Hermitian with respect to $\eta^{0}$. Therefore, we can conclude that the operator $H$ of the Hamiltonian form with electromagnetic interaction is neither Hermitian in the standard sense (for the sake of the anomalous term) nor with respect to $\beta^{0}$. It does not matter, $H$ should be Hermitian with respect to $\eta^{0}$ in order to provide real eigenvalues.

\section{Four-current conserved}

\noindent Returning to the ambiguity with the electromagnetic coupling, let us begin with the equation of motion. The conservation law for $J^{\mu }$
follows from the standard procedure of multiplying (\ref{dkp2}) and its
complex conjugate by $\bar{\psi}$ from the left and by $\eta ^{0}\psi $ from
the right, respectively. On the other hand, by carrying through calculations
similar to those ones using the DKP equation, the Schr\"{o}dinger-like
equation (\ref{sch}) leads to%
\begin{eqnarray}
\partial _{\mu }J^{\mu } &=&\left[ \left( D_{j}\right) ^{\ast }\bar{\psi}%
\right] \beta ^{0}\beta ^{0}\beta ^{j}\psi +\bar{\psi}\beta ^{j}\beta
^{0}\beta ^{0}D_{j}\psi   \notag \\
&&  \label{DJ} \\
&=&\left( \partial _{j}\bar{\psi}\right) \beta ^{0}\beta ^{0}\beta ^{j}\psi +%
\bar{\psi}\beta ^{j}\beta ^{0}\beta ^{0}\left( \partial _{j}\psi \right)
+ieA_{j}\bar{\psi}[\beta ^{j},\beta ^{0}\beta ^{0}]\psi   \notag
\end{eqnarray}%

\noindent In this case one sees that the malediction of a source term falls on $J^{\mu
}$.

Up to this point the physical components of the DKP spinor have not come
into the story at all. The contradictory results involving the source terms
can be solved by following the prescription put forward in Refs. \cite{now}
and \cite{lun}. Instead of working with a specific representation for the
matrices $\beta ^{\mu }$ we choose an alternative way.

\subsection{Scalar sector}

To select the physical component of the DKP field for the scalar sector (spin-0 sector), we define the operator~\cite{ume}
\begin{equation}\label{opep}
    P=-(\beta^{0})^2(\beta^{1})^{2}(\beta^{2})^{2}(\beta^{3})^{2}\,,
\end{equation}

\noindent which satisfies $P^{2}=P$, $P^{\mu}=P\beta^{\mu}$ and $^{\nu}\!P=(P^{\nu})^{\dagger}=\beta^{\nu}P$. As it is shown in \cite{ume}, $P\psi$ and $P^{\mu}\psi$ transform as a (pseudo)scalar and a (pseudo)vector under an infinitesimal Lorentz transformation, respectively.

The spin-0 sector can be expressed by the set of operators $\left\{ P,\,^{\mu }\!P,P^{\mu
},\,^{\mu }\!P^{\nu }\right\} $ with the properties \cite{FIS}:%
\begin{eqnarray}
P\left( P^{\mu }\right)  &=&P^{\mu },\quad \left( ^{\mu }\!P\right)
P=\,^{\mu }\!P\quad   \notag \\
&&  \notag \\
\left( P^{\mu }\right) P &=&P\left( ^{\mu }\!P\right) =0  \label{P} \\
&&  \notag \\
\left( ^{\mu }\!P\right) \left( P^{\nu }\right)  &=&\,^{\mu }\!P^{\nu
},\quad \left( P^{\mu }\right) \left( ^{\nu }\!P\right) =g^{\mu \nu }P\quad
\notag
\end{eqnarray}%

\noindent Hence%
\begin{eqnarray}
P\left( \,^{\mu }\!P^{\nu }\right) &=&\left( \,^{\mu
}\!P^{\nu }\right) P=0,\quad \left( P^{\mu }\right) \left( P^{\nu }\right)
=\left( ^{\nu }\!P\right) \left( ^{\mu }\!P\right) =0  \notag \\
&&  \label{p} \\
\beta ^{\mu } &=&P^{\mu }+\,^{\mu }\!P,\quad \bar{\psi}P=\left( P\psi
\right) ^{\dagger }  \notag
\end{eqnarray}%

\noindent in such a way that the DKP equation becomes%
\begin{equation}
D_{\mu }\left( P^{\mu }\psi \right) =-im\left( P\psi \right) ,\quad D^{\mu
}\left( P\psi \right) =-im\left( P^{\mu }\psi \right)   \label{eq}
\end{equation}%

\noindent which provides%
\begin{equation}
\left( D^{\mu }D_{\mu }+m^{2}\right) \left( P\psi \right) =0,\quad \left(
D^{\mu }D_{\mu }+m^{2}\right) \left( P^{\nu }\psi \right) =0  \label{kg}
\end{equation}%

\noindent These results tell us that all elements of the column matrices $P\psi $ and $P^{\mu }\psi $ obey the
KG equation with minimal coupling and that $P^{\mu }\psi $ is expressed in terms of the
covariant derivative of $P\psi $. Then, acting $P$ upon the spinor DKP $\psi$ selects the scalar sector of DKP theory, making explicitly clear that it describes a spin-0 particle embedded in a electromagnetic field. Following this innovative view of the DKP
spinor, Ref. \cite{lun} shows that the redundant components of $\psi $ are
projected out, $\psi $ and $P\psi $ are both compatible with gauge
invariance and the anomalous term in the Hamiltonian version has no physical
consequence. Now, return our attention to the DKP current. The $P$-algebra implies that%
\begin{eqnarray}
J^{\mu } &=&\frac{1}{2}\,\bar{\psi}\left( P^{\mu }+\,^{\mu }\!P\right) \psi   \notag \\
&&  \notag \\
&=&\frac{i}{2m}\left\{ \left( P\psi \right) ^{\dagger }\left[ D^{\mu }\left(
P\psi \right) \right] -\left[ \left( D^{\mu }\right) ^{\ast }\left( P\psi
\right) ^{\dagger }\right] \left( P\psi \right) \right\}   \label{cur} \\
&&  \notag \\
&=&\frac{i}{2m}\left\{ \left( P\psi \right) ^{\dagger }\left[ \partial ^{\mu
}\left( P\psi \right) \right] -\left[ \partial ^{\mu }\left( P\psi \right)
^{\dagger }\right] \left( P\psi \right) \right\} -\frac{e}{m}A^{\mu }\left(
P\psi \right) ^{\dagger }\left( P\psi \right)   \notag
\end{eqnarray}%

\noindent This is nothing but the KG current. In other words, \noindent\ the
DKP current is equivalent to the KG current. Hence, $\partial
_{\mu }J^{\mu }=0$, as derived from the DKP equation, can be seen as a
natural result. Indeed, it follows from the $P$-algebra that%
\begin{equation}
\beta ^{0}\beta ^{0}\beta ^{j}=P^{j},\quad \beta ^{j}\beta ^{0}\beta
^{0}=\,^{j}\!P  \label{be}
\end{equation}%

\noindent Thus (\ref{DJ}) can be written as
\begin{eqnarray}
\partial _{\mu }J^{\mu } &=&\left[ \left( D_{j}\right) ^{\ast }\bar{\psi}%
\right] P^{j}\psi +\left[ \bar{\psi}\left( ^{j}\!P\right) \right] D_{j}\psi
\notag \\
&&  \notag \\
&=&\frac{i}{m}\left\{ \left[ \left( D_{j}\right) ^{\ast }\left( P\psi
\right) ^{\dagger }\right] D^{j}\left( P\psi \right) -\left[ \left(
D^{j}\right) ^{\ast }\left( P\psi \right) ^{\dagger }\right] D_{j}\left(
P\psi \right) \right\}  \\
&&  \notag \\
&=&0  \notag
\end{eqnarray}%

\subsection{Vectorial sector}

Now we discuss the vector sector (spin-1 sector) of the DKP theory. Similarly to the scalar sector, we can select the physical components of the DKP field for the spin-1 sector, so we define the operator \cite{ume}
\begin{equation}\label{oper}
    R^{\mu}=(\beta^{1})^{2}(\beta^{2})^{2}(\beta^{3})^{2}\left[ \beta^{\mu}\beta^{0}-g^{\mu0} \right]
\end{equation}

\noindent which satisfies $R^{\mu\nu}=R^{\mu}\beta^{\nu}$ and $R^{\mu\nu}=-R^{\nu\mu}$. As it is shown in \cite{ume}, $R^{\mu}\psi$ and $R^{\mu\nu}\psi$ transform as a (pseudo)vector and (pseudo)tensor quantities under an infinitesimal Lorentz transformation, respectively.

The spin-1 sector can be expressed by the set of operators
$\{^{\mu }\!\,V^{\nu },\,^{\mu }\!\,V^{\nu \lambda },\,^{\nu
\lambda }\!\,V^{\mu },\,^{\nu \lambda }\!\,V^{\mu \sigma }\}$ \cite{FIS},
with%
\begin{eqnarray}
^{\mu }\!\,V^{\nu } &=&\left( ^{\mu }\!R\right) \left( \!R^{\nu }\right)
,\quad ^{\mu }\!\,V^{\nu \lambda }=\left( ^{\mu }\!R\right) \left( \!R^{\nu
\lambda }\right)   \notag \\
&&  \label{v1} \\
^{\nu \lambda }\!\,V^{\mu } &=&\left( ^{\nu \lambda }\!R\right) \left(
\!R^{\mu }\right) ,\quad ^{\nu \lambda }\!\,V^{\mu \sigma }=\left( ^{\nu
\lambda }\!R\right) \left( \!R^{\mu \sigma }\right)   \notag
\end{eqnarray}%

\noindent where%
\begin{eqnarray}
\left( \!R^{\mu }\right) \left( ^{\nu }\!R\right)  &=&\left( \!R^{0}\right)
g^{\mu \nu },\quad \left( \!R^{\mu }\right) \left( \!R^{\nu \lambda }\right)
=\left( \!R^{\nu \lambda }\right) g^{\mu 0},\quad \left( \!R^{\mu }\right)
\left( \!R^{\nu }\right) =\left( \!R^{\nu }\right) g^{\mu 0}  \notag \\
&&  \notag \\
\left( \!R^{\mu \nu }\right) \left( ^{\lambda }\!R\right)  &=&\left(
\!R^{\mu }\right) \left( ^{\nu \lambda }\!R\right) =\left( \!R^{\mu \nu
}\right) \left( \!R^{\lambda }\right) =0  \label{v2} \\
&&  \notag \\
\left( \!R^{\mu \nu }\right) \left( ^{\lambda \sigma }\!R\right)  &=&\left(
\!R^{0}\right) \Delta ^{\mu \nu \lambda \sigma },\quad \Delta ^{\mu \nu
\lambda \sigma }=g^{\mu \sigma }g^{\nu \lambda }-g^{\mu \lambda }g^{\nu
\sigma }  \notag
\end{eqnarray}%

\noindent In view \ of (\ref{v2}) one has%
\begin{eqnarray}
\left( ^{\mu }\!\,V^{\nu \lambda }\right) \left( ^{\rho \sigma }\!\,V^{\tau
}\right)  &=&\left( ^{\mu }\!\,V^{\tau }\right) \Delta ^{\nu \lambda \rho
\sigma }  \notag \\
&&  \notag \\
\left( ^{\mu }\!\,V^{\nu \lambda }\right) \left( ^{\tau }\!\,V^{\rho \sigma
}\right)  &=&\left( ^{\nu \lambda }\!\,V^{\mu }\right) \left( ^{\rho \sigma
}\!\,V^{\tau }\right) =0  \notag \\
&&  \label{v30} \\
\beta ^{\mu } &=&\sum_{\lambda }\left( ^{\mu \lambda }\!\,V^{\lambda
}+\,^{\lambda }\!\,V^{\lambda \mu }\right)   \notag \\
&&  \notag \\
\bar{\psi}\left( \!R^{0}\right)  &=&\left( \!R^{0}\psi \right) ^{\dagger
}\eta ^{00},\quad \bar{\psi}\!\left( ^{i0}\!R\right) =\left( \!R^{i0}\psi
\right) ^{\dagger }\eta ^{00}  \notag
\end{eqnarray}%

\noindent in such a way that the DKP equation becomes%
\begin{eqnarray}
D_{\mu }\left( R^{\nu \mu }\psi \right)  &=&-im\left( R^{\nu }\psi \right)
,\quad \left( R^{\mu \nu }\psi \right) =-\frac{i}{m}U^{\mu \nu }  \notag \\
&&  \label{v4} \\
U^{\mu \nu } &=&D^{\mu }\left( R^{\nu }\psi \right) -D^{\nu }\left( R^{\mu
}\psi \right)   \notag
\end{eqnarray}%

\noindent which leads to%
\begin{eqnarray}
D_{\mu }U^{\mu \nu }+m^{2}\left( R^{\nu }\psi \right)  &=&0  \notag \\
&&  \label{v6} \\
D_{\mu }\left( R^{\mu }\psi \right) &=&\frac{ie}{2m^{2}}F_{\mu \nu }U^{\mu \nu
}  \notag
\end{eqnarray}%

\noindent These results tell us that all elements of the column matrix $R^{\mu}\psi$ obey the Proca equation interacting minimally with an electromagnetic field. So, similarly to the scalar sector, this procedure selects the vector sector of DKP theory, making explicitly clear that it describes a spin-1 particle embedded in a electromagnetic field. A little calculation shows that
\begin{eqnarray}
J^{\mu } &=&\frac{1}{2}\sum_{\lambda }\bar{\psi}\left( ^{\mu \lambda }\!\,V^{\lambda
}+\,^{\lambda }\!\,V^{\lambda \mu }\right) \psi   \notag \\
&&  \notag \\
&=&\frac{1}{2}\sum_{\lambda }\bar{\psi}\left( ^{\lambda }\!R\right) \!\left( R^{\lambda
\mu }\psi \right) +\frac{1}{2}\sum_{\lambda }\bar{\psi}\left( ^{\mu \lambda }\!R\right)
\left( \!R^{\lambda }\psi \right)   \label{v7} \\
&&  \notag \\
&=&-\frac{i}{2m}\sum_{\lambda }\left[ \left( \!R^{\lambda }\psi \right)
\left( U^{\mu \lambda }\right) ^{\ast }-\bar{\psi}\left( ^{\lambda
}\!R\right) U^{\mu \lambda }\right]   \notag
\end{eqnarray}%

\noindent Which shows that $J^{\mu }$ is completely equivalent to the Proca current.
Moreover, in order to evaluate $\partial _{\mu }J^{\mu }$ we use the
relations
\begin{eqnarray}
\beta ^{0}\beta ^{0}\beta ^{i} &=&\sum_{\lambda }\left( ^{\lambda
}\!R\right) \left( \!R^{\lambda i}\right) -\left( ^{0}\!R\right) \left(
\!R^{0i}\right) -\left( ^{0i}\!R\right) \left( \!R^{0}\right)   \notag \\
&&  \label{v8} \\
\beta ^{i}\beta ^{0}\beta ^{0} &=&\sum_{\lambda }\left( \!R^{i\lambda
}\right) \left( R^{\lambda }\right) -\left( ^{i0}\!R\right) \left(
\!R^{0}\right) -\left( ^{i}\!R\right) \left( \!R^{i0}\right)   \notag
\end{eqnarray}%

\noindent so that (\ref{DJ}) results in
\begin{eqnarray}
\partial _{\mu }J^{\mu } &=&\bar{\psi}\left( \!R^{0}\right) D_{i}\!\left(
R^{0i}\psi \right) +\left[ D_{i}^{\ast }\bar{\psi}\left( ^{i0}\!R\right) %
\right] \!\left( R^{0}\psi \right)   \notag \\
&&  \notag \\
&=&-\left( \!R^{0}\psi \right) ^{\dagger }D_{i}\!\left( R^{0i}\psi \right) -
\left[ D_{i}^{\ast }\left( \!R^{i0}\psi \right) ^{\dagger }\right] \!\left(
R^{0}\psi \right)   \label{v10} \\
&&  \notag \\
&=&0  \notag
\end{eqnarray}%

\noindent Again, the correct physical components of the DKP spinor makes $J^{\mu}$ conserved if one uses either the equation of motion or the Hamiltonian. Therefore, there is no problem in introducing the minimal coupling in the equation of motion or in the form Hamiltonian, because these two ways provide a conserved four-current. Recently, the projectors $P$, $P^{\mu}$, $R^{\mu}$ and $R^{\mu\nu}$ have satisfactorily been used to found analytical solutions for spin-0 and spin-1 particles \cite{esdras,ben3}.

\section{Final Remarks}

In summary, using the conservation of the four-current and the correct interpretation of the physical components of the DKP spinor, we tried to clarify the ambiguity of the electromagnetic coupling in the DKP theory. From this point of view, the ambiguity seen by Kemmer in his original work \cite{kem} does not exist, because the current is conserved in both versions of the DKP theory if one uses the correct physical components of the DKP spinor as prescribed in \cite{lun}. Therefore, it does not matter if one either put the electromagnetic coupling straight in the Hamiltonian or in the equation of motion. Furthermore, $J^{\mu }$ reduces to the KG current or to the Proca current when one selects the appropriate sector of the theory, as should be expected from equivalent theories. Thanks to the algebraic properties of the projectors developed in \cite{FIS} neither representation for the DKP matrices was used for reaching this conclusion, even if the physical fields depend on the explicit representation of the DKP matrices. Also, we analyzed the Hermiticity of the Hamiltonian and we showed that the operator $H$ of the Hamiltonian form is neither Hermitian in the usual sense nor with respect to $\beta^{0}$ as argued by Zeleny \cite{zel}. As a matter of fact, the operator $H$ should be Hermitian with respect to $\eta^{0}$, $\left[ \eta^{0}\left( \beta^{0}H \right)\right]^{\dagger}=\eta^{0}\left( \beta^{0}H \right)$, in order to provide real eigenvalues. Finally, our results corroborate and complement the results presented in \cite{now,lun} and also they shed some light on some widespread misconceptions about the Hermiticity of the Hamiltonian form in the DKP theory.

\begin{acknowledgments}
The authors are indebted to the anonymous referee for an excellent and constructive review. This work was supported in part by means of funds provided by CAPES and CNPq. This work was partially done during a visit (L. B. Castro) to UNESP - Campus de Guaratin\-gue\-t\'{a}.
\end{acknowledgments}

\end{document}